\documentclass[prb,onecolumn,nofootinbib,citeautoscript,10pt,notitlepage, showkeys]{revtex4-2}
\pdfoutput=1

\usepackage{float}
\usepackage{array}
\usepackage{slashed,bbold}

\usepackage{amsmath,amssymb,bm} 
\usepackage{graphicx}

\usepackage[tight]{subfigure} 

\usepackage{color} 
\usepackage[papersize={8.5in,11in}]{geometry}

\usepackage{color}
\definecolor{darkblue}{rgb}{0.,0.,0.4}
\definecolor{darkred}{rgb}{0.5,0.,0.}
\definecolor{BlueViolet}{RGB}{138,43,226}
\definecolor{SkyBlue}{RGB}{30,144,255}
\definecolor{DarkGreen}{RGB}{0,100,0}
\usepackage[pdftex,colorlinks=true,linkcolor=darkblue,citecolor=blue,urlcolor=darkred]{hyperref}

\def \nn{\nonumber \\}

\begin{document}

\title{Nature of Andreev bound states in Josephson junctions of triple-point semimetals}

\author{Ipsita Mandal}
\email{ipsita.mandal@snu.edu.in}

\affiliation{Department of Physics, Shiv Nadar Institution of Eminence (SNIoE), Gautam Buddha Nagar, Uttar Pradesh 201314, India
\\and\\
Freiburg Institute for Advanced Studies (FRIAS), University of Freiburg, D-79104 Freiburg, Germany}

\begin{abstract}
We study superconductor-barrier-superconductor (S-B-S) Josephson junctions constructed out of two-dimensional and three-dimensional triple-point semimetals, which feature a threefold degeneracy at a single nodal point. We assume a weak and homogeneous s-wave pairing in each superconducting region, and a potential difference is applied across a piece of normal-state semimetal to create the barrier region. We compute the wavefunctions of the Andreev bound states (ABSs), considering the thin-barrier limit. The appropriate boundary conditions at the S-B and B-S junctions allow us to compute the discrete energy eigenvalues $\pm |\varepsilon| $ of the ABSs. We get two distinct solutions for $|\varepsilon| $. This result differs from that in graphene and Weyl semimetals, where one obtains only one solution for $|\varepsilon| $. The multifold nature of the triple-point fermions is responsible for this difference. We also illustrate the behaviour of the Josephson current flowing across the S-B-S junction.
\end{abstract}


\maketitle


\section{Introduction}
\begin{figure*}[t]
\subfigure[]{\includegraphics[width = 0.22 \textwidth]{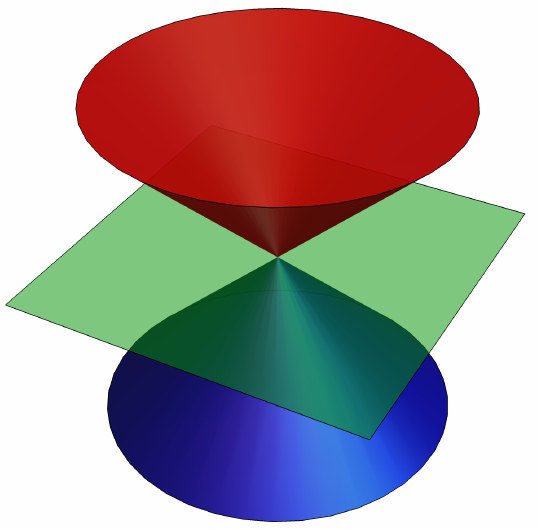}}
\hspace{2 cm}
\subfigure[]{\includegraphics[width = 0.5 \textwidth]{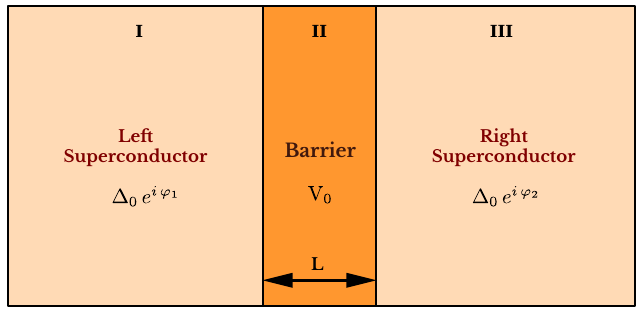}}
\caption{\label{figsetup}
Schematics of the (a) dispersion aound a triply-degenerate nodal point; (b) S-B-S junction configuration. For the pseudospin-1 semimetal, two of the bands demonstrate an isotropic linear-in-momentum dispersion behaviour, while the third one is a flat band with momentum-independent constant energy.
}
\end{figure*}

Since the last decade, there have been intensive studies of bandstructures harbouring symmetry-protected band-crossing points in 
the Brillouin zone (BZ) \cite{burkov11_weyl, yan17_topological,  bernevig, bernevig2, lv}, giving rise to semimetals. The low-energy  $\mathbf k \cdot \mathbf p$ Hamiltonian near the band-crossings gives rise to quasiparticles carrying pseudospin values equal to $ \varsigma  $, when $(2\, \varsigma + 1) $ bands touch at the degeneracy point. We note that, whereas in high-energy physics, a relativistic electron is described by a spin-1/2 fermion (with the intrinsic spin quantum number fixed by Lorentz invariance), the pseudospin of an emergent electronic excitation (still having the spin-1/2 value) in nonrelativistic condensed matter physics can be different from 1/2. In fact, for each of the 230 space groups, the pseudospin quantum numbers are dictated by the irreducible representations of the little group of lattice symmetries at the high-symmetry points in the BZ \cite{bernevig}. The dimensionality of the irreducible representation corresponds to the number of bands crossing at the high-symmetry point. In particular, a threefold degeneracy can arise in a crystal lattice when the nodal point is protected by (I) nonsymmorphic symmetries \cite{bernevig} or (II) symmorphic rotation, combined with mirror symmetries \cite{spin13d1, spin13d2, spin13d4, ady-spin1}. A schematic illustration of the triply-degenerate nodal point is shown in Fig.~\ref{figsetup}(a).


In this paper, we focus on triple-point fermionic quasiparticles carrying pseudospin value $ \varsigma = 1$, whose effective low-energy continuum Hamiltonian is of the form $\mathbf{k} \cdot \boldsymbol{\mathcal{S}}$, where $ \boldsymbol{\mathcal{S}}$ represents the vector of the three matrices which form the spin-1 representation of the SO(3) group.
This leads to the emergence of two-dimensional (2d) and three-dimensional (3d) semimetals with pseudospin-1 quasiparticles \cite{optical_lat1, optical_lat2, cold-atom, lv, spin12d1, spin12d2, spin12d3, spin13d1, spin13d2, spin13d3, spin13d4, shen,lan, urban, peng-he, lai, ips3by2, krish-spin1, ips-cd}, sometimes dubbed as ``Maxwell fermions'' \cite{cold-atom}, since their pseudospin quantum number is analogous to the spin-1 quantum number of the photons (which are described by the Maxwell equations). The 3d versions with linear and isotropic dispersion are the straightforward generalizations of the Weyl semimetal Hamiltonian $\mathbf{k} \cdot \boldsymbol{\sigma}$, with $\boldsymbol{\sigma}$ representing the vector operator consisting of the three Pauli matrices, thus representing the spin-1/2 representation of the SO(3) group. One can show that the threefold degeneracies carry nonnzero Chern numbers (equalling $\pm 2$) \cite{ bernevig, igor, ips-cd, ips-magnus}, analogous to the pseudospin-3/2 Rarita-Schwinger-Weyl (RSW) semimetals~\cite{long, igor, igor2, isobe-fu, zhu, fang, ips3by2, ips-cd, ips-magnus, ips_jns, ips_jj_rsw} harbouring fourfold-degenerate nodal points, thus reflecting the nontrivial topological character of the corresponding bandstructures. Here, we continue the ongoing efforts to unravel the distinct signatures of the triple-point quasiparticles via various transport properties \cite{shen,lan, ips3by2, krish-spin1, ips-cd}. In particular, the transport property under consideration will be the Josephson current $I_J$ \cite{josephson, likharev, waldram}, arising in
configurations consisting of junctions between the normal (abbreviated by ``N'') and the superconducting (abbreviated by ``S'') phases of 2d and 3d pseudospin-1 semimetals.

When two superconducting regions are coupled to each other by a weak link between them, $I_J$ is the equilibrium dissipationless current flowing across the junction. Since the Andreev surface states of the two superconductors hybridize to form Andreev bound states (ABSs) at the junction, they contribute to $I_J$.
Here, we extend the study of such Josephson effects to 2d and 3d semimetals \cite{titov-graphene, bolmatov_graphene_sns, krish-moitri, emil_jj_WSM, debabrata-krish, debabrata, ips_jj_rsw}, where the superconducting regions consist of homogeneous s-wave pairing, and the weak link is created by introducing a tunneling barrier.
One can consider two alternative arrangements: (I) S-N-S junctions~\cite{titov-graphene,bolmatov_graphene_sns}; (II) S-B-S (where ``B'' indicates a rectangular potential barrier in the N region) junctions~\cite{krish-moitri, emil_jj_WSM, debabrata-krish, debabrata, ips_jj_rsw, ips-abs-semid}.
In experiments, superconductivity can be induced in the appropriate region via proximity-effect by placing a conventional s-wave superconductor on top of an electrode made of the semimetal \cite{proximity-sc}, and the barrier region can be created by applying a gate voltage $V_0$ across N. The schematics of an S-B-S set-up is illustrated in Fig.~\ref{figsetup}(b), where we demarcate the left superconducting region as ``region I'', the middle barrier region as ``region II'', and the right superconducting region as ``region III''.
We will focus on the short-barrier regime, which implies that the barrier thickness $ L $ along the propagation direction is taken to be $ L \ll \xi $, where $\xi $ is the superconducting coherence length. In other words, $ \xi$ denotes the length scale over which the probability of the wavefunctions of the ABSs decay inside the superconductor, as we move away from the junction location.

Let us set up some notations to characterize the S-B-S configuration. We denote the complex superconducting order parameter by $\Delta = \Delta_0 \, e^{i\,\varphi} $, with $\varphi$ representing the phase variable. Denoting the energy of the eigenstates by $\varepsilon$, a set of discrete states is obtained for $ |\varepsilon| <\Delta_0 $, which comprises the ABSs referred to in the discussions above. They are also known as the subgap excitations because their energy values lie within the superconducting gap region. The eigenstates with $|\varepsilon | >\Delta_0 $ form a continuum, which do not decay within the bulk of the superconductors. Because three bands cross at a single nodal point of a triple-point semimetal, the nature of the ABSs must be different from those arising in the Dirac \cite{titov-graphene,krish-moitri}, semi-Dirac \cite{ips-abs-semid}, Weyl \cite{debabrata-krish}, multi-Weyl \cite{debabrata-krish, debabrata}, and RSW \cite{ips_jj_rsw} semimetals. 
In particular, for doubly-degenerate nodal points, when the propagation direction is along an axis where dispersion is linear-in-momentum, it has been found that \cite{titov-graphene,krish-moitri,debabrata-krish} that the energy of the ABSs in the thin-barrier-limit is given by $ \varepsilon = \pm \, \Delta_0 \, \sqrt{1 - T_N\, \sin^2 \left( \varphi_{12} / 2\right) }$. Here, $\varphi_{12} $ is the difference of the superconducting phases on the two sides of the barrier region [cf. Fig.~\ref{figsetup}(b)], and $T_N$ is the transmission coefficient in an analogous set-up with the two superconducting regions replaced by the normal state of the semimetal. This simple relation arises from the fact that the solution for $\beta \equiv \varepsilon/\Delta_0 $ is obtained from a polynomial equation of the form $ \tilde{\mathcal{B} } \cos(2\beta) + \tilde{\mathcal{ C }} \cos \varphi_{12} =0 $ \cite{krish-moitri}, where $ \tilde{\mathcal{B} }$ and $\tilde {\mathcal{C}}$ are purely functions of the barrier strength $V_0$, the Fermi energy $E_F$, and the magnitude of the perpendicular-to-propagation momentum-component, $k_\perp $. However, such a simple equation does not arise for generic cases, as exemplified by earlier studies on semi-Dirac \cite{ips-abs-semid} and RSW \cite{ips_jj_rsw} semimetals.

We consider the propagation of quasiparticles and quasiholes in a slab with a side-length $W$, where $W$ is assumed to be large enough to impose periodic boundary conditions. The slab is 2d (3d) when we consider a 2d (3d) pseudospin-1 semimetal. As pointed out earlier, we will compute the energy values of the ABSs in the thin-barrier-limit (also known as the short-barrier regime), which is quantified by the relations $V_0 \rightarrow  \infty $ and $L \rightarrow 0 $, with $\chi \equiv V_0\, L$ held fixed at a finite value. In this limit, the ABSs are found to be the dominant contributors to the total Josephson current $I_J$ \cite{been_houten, titov-graphene, zagoskin, tanaka_review}, because the contributions from the excited states in the continuum are smaller by a factor of $L/\xi$. 

The paper is organized as follows. In Sec.~\ref{sec2d}, we consider the 2d triple-point fermions and derive the discrete spectrum for the ABSs resulting from an S-B-S configuration. Therein, we explain how the S-B-S junction is constructed, for theoretical analysis, via the electron-like and hole-like wavefunctions and the Bogoliubov–de Gennes (BdG) Hamiltonian. This is followed by Sec.~\ref{sec3d}, where we repeat the same exercise for the 3d version of the triple-point fermions. Finally, we end with a summary and an outlook in Sec.~\ref{secsum}.

\section{2d model}
\label{sec2d}

The 2d pseudospin-1 quasiparticles are the generalizations of the pseudospin-1/2 Dirac quasiparticles in graphene at half-filling. They can be realized in 2d tight-binding models for cold fermionic atoms in square optical lattices \cite{cold-atom} and, via first principle calculations, have been predicted to exist in materials like phosphorene oxide \cite{spin12d1},
hexacoordinated Mg$_2$C monolayer \cite{spin12d2}, and monolayer dialkali-metal monoxides (e.g., Na$_2$O and K$_2$O) \cite{spin12d3}. Working in natural units (with $\hbar $ set to unity), the effective low-energy continuum
Hamiltonian, in the vicinity of a nodal point, is given by
\begin{align}
\label{eqham2d}
\mathcal{H}_{2d}(\mathbf  k) = v_F \left(k_x \,\mathcal S_x + k_y \,\mathcal S_y \right).
\end{align}
Here, $\mathbf k = k_x \, \hat {\mathbf x} + k_y \, \hat {\mathbf y} $, $ \boldsymbol{\mathcal S} $ represents the vector spin-1 operator with the three components
\begin{align}
\mathcal{S}_x = \frac {1} {\sqrt{2}}
\begin{pmatrix}
0&1&0\\1&0&1\\0&1&0
\end{pmatrix} ,\quad
\mathcal{S}_y =\frac{1}{\sqrt{2}}
\begin{pmatrix}
0&- i &0\\ \mathrm{i} &0&-  i \\
0&  i  &0
\end{pmatrix} ,\quad
\mathcal{S}_z =
\begin{pmatrix}
1&0&0\\0&0&0\\0&0&-1
\end{pmatrix},
\end{align}
and $v_F$ denotes the magnitude of the Fermi velocity. Henceforth, we will set $v_F = 1$ for the sake of simplifying the notations.

The energy eigenvalues are given by
\begin{align}
\label{eqeigen}
\varepsilon_{s}(\mathbf{k}) = s  \, k \text{ and }
\varepsilon_{0}(\mathbf k) = 0\,,
\end{align}
where $k= |\mathbf k|$ and $s \in \lbrace + , \, -\rbrace $, showing that two linearly-dispersing bands and a nondispersive flat band cross at $k=0$ [cf. Fig.~\ref{figsetup}(a)]. Clearly, the ``$+$" and ``$-$" signs refer to the linearly-dispersing conduction and valence bands, respectively.
A set of orthogonal eigenvectors is captured by
\begin{align}
\label{eqwv1}
\Psi_s (\mathbf k) & = \left(\frac {k} {k_x + i \, k_y}  \qquad  \sqrt {2}  \,s  \qquad
 \frac {k} {k_x - i  \,k_y} \right )^T \left(\text{for energy} = s\, k \right)
\nn 
 \nn \text{ and }
  \Psi_0 (\mathbf k) & = \left ( -k_x + i \, k_y \qquad 0
 \qquad  k_x + i \, k_y \right )^T \left(\text{for the flat band}\right).
\end{align}

The probability-current-density operator for this system is captured by $ \check{ \mathbf{j} }_{2d} (\mathbf k)= \nabla_{\mathbf{k}} \mathcal{H}_{2d}(\mathbf  k)
= \mathcal{S}_x \, \hat {\mathbf x} + \mathcal{S}_y \, \hat {\mathbf y}$.
Therefore, for a wavefunction $\psi \equiv  \left( c_1  \quad c_2  \quad c_3 \right ) ^T $, the expectation value of the probability flux is given by
\begin{align}
\label{eqcur1}
\psi^\dagger \, \check{ \mathbf{j} }_{2d} \,\psi =
\sqrt 2 \, {\text{Re} \left[ c_ 2^* \left ( c_ 1 + c_ 3 \right) \right]} 
\,  \hat {\mathbf x}
 + 
\sqrt 2 \, {\text{Im} \left[ c_ 2^* \left ( c_ 1 - c_ 3 \right) \right]} 
\,  \hat {\mathbf y}\,.
\end{align}
This immediately tells us that the contribution of the flat-band wavefunction to the local probability-current density is
$
 \Psi_0^\dagger \,\check{ \mathbf{j}}_{2d} \, \Psi_0 = \mathbf 0\,,$ showing that it does not contribute to the determination of the boundary conditions \cite{lai, ips3by2}.\footnote{See the appendix of Ref.~\cite{ips_jns} for more details. Although we have the question of propagation of particles only in Ref.~\cite{ips_jns}, and not holes, the discussion equally applies here where we need to consider both the particle- and hole-wavefunctions.} Hence, we need to consider only $\Psi_\pm$ for imposing the appropriate boundary conditions for a system comprising the normal state.
Eq.~\eqref{eqcur1} also tells us that if the quasiparticles encounter a perpendicular barrier, say at $x = x_0$, while propagating along the $x$-direction, the conservation of the probability-current density leads to the two boundary conditions \cite{urban, fang},
\begin{align}
\label{eqbdy1}
\lim_{ \delta \rightarrow 0 }
\left[c_1(x_0-\delta, y ) + c_3 (x_0-\delta,y ) \right ] = \lim_{ \delta \rightarrow 0 }
 \left [ c_1(x_0 + \delta,y ) + c_3 (x_0 +\delta,y ) \right ]
 \text{ and }
\lim_{ \delta \rightarrow 0 }  c_2(x_0-\delta,y ) = 
\lim_{ \delta \rightarrow 0 } c_2 (x_0 + \delta, y ) \,.
\end{align}
If the Fermi energy cuts the dispersion profile at an energy value $E$, then for propagation along the $x$-direction, the wavefunction will have the exponential factor $e^{i\, \text{sgn}(E) \,k_x \, x}$, where $ k_x  = \sqrt{ E^2-k_\perp^2 }$ and $k_\perp = | k_y| $.

\subsection{S-B-S junction's interfaces oriented perpendicular to the $x$-direction}
\label{secsbs2d}

Here we consider the propagation of the quasiparticles/quasiholes along the $x$-axis. We note that the final results would be the same if we had taken the propagation direction to be along the $y$-axis, which follows from the isotropic nature of the dispersion.

Setting up the S-B-S configuration as shown in Fig.~\ref{figsetup}(b)], we model the superconducting pair potential as
\begin{align}
\label{eqscorder0}
\Delta (x) =\begin{cases} 
\Delta_0\,e^{i\,\varphi_1 }\,{\mathcal S}_0 
 &\text{ for }  x \leq 0   \\
0 &\text{ for } 0 < x < L \\
\Delta_0\,e^{i\,\varphi_2 } \,{\mathcal S}_0  &\text{ for } x \geq L
\end{cases}, \quad
{\mathcal S}_0 = {\mathbb{1}}_{3 \times 3} \,,
\end{align}
representing Cooper pairing in the s-wave channel. Due to the presence of the barrier region, we need to consider the potential energy
\begin{align}
V(x)
=\begin{cases} 
0
 &\text{ for }  x \leq 0 \text{ and } x \geq L  \\
 V_0 &\text{ for } 0 < x < L 
\end{cases}.
\end{align} 

The resulting BdG Hamiltonian is given by
\begin{align}
& H(\mathbf r) = \frac{1} {2} \sum_{\mathbf k} \Psi^\dagger_{\mathbf k} \,H_{\text{BdG}} (\mathbf k)
\Psi_{\mathbf k} , \quad
\Psi_{\mathbf k} = \begin{pmatrix}
c_1 (\mathbf k) & c_2 (\mathbf k) & c_3 (\mathbf k)  & 
c_1^\dagger (-\mathbf k) & c_2 ^\dagger(-\mathbf k) & 
c_3 ^\dagger(-\mathbf k) 
\end{pmatrix}^T\,, \nn
& H_{\text{BdG}} (\mathbf k, \mathbf r) =
\begin{pmatrix}
\mathcal{H}_{2d}(\mathbf  k) -E_F + V(x)  & \Delta(x)  \\ 
 \Delta^\dagger(x) &  E_F- V(x) -\mathcal{H}_{2d}^T(-\mathbf  k, \mathbf r)  \\  
\end{pmatrix},
\label{eq_bdg0}
\end{align}
where the subscripts $\lbrace 1, 2, 3\rbrace $ on the fermionic creation and annihilation operators represent the three distinct band indices. We will work assuming the energy-scale hierarchies embodied by $V_0 \gg E_F  \gg \Delta_0 $ and $(V_0-E_F) \gg E_F $.
While the condition $\Delta_0 \ll E_F$ ensures that the mean-field approximation, applicable for using the BdG formalism, is valid, the second condition $ (V_0-E_F) \gg E_F $ results from taking the thin-barrier limit.

The electron-like and the hole-like BdG quasiparticles are obtained from the eigenvalue equation
\begin{align}
H_{\text{BdG}} ( \mathbf k\rightarrow -i \boldsymbol{\nabla}_{\mathbf r},
\mathbf r) 
\,\psi_{\mathbf k} (\mathbf r) 
= \varepsilon \, \psi_{\mathbf k} (\mathbf r) \,,
\end{align}
where $\mathbf r=  x \, \hat {\mathbf x} + y \, \hat {\mathbf y}$ is the 2d position vector. If $ \psi_N (\mathbf k)$ is an eigenfunction of $\mathcal{H}_{2d}(\mathbf  k)$ (with the superconducting phase factor of $\varphi $), then the electron-like and hole-like eigenfunctions of $ H_{\text{BdG}} ( \mathbf k) $ are given by the expressions~\cite{timm}
\begin{align}
\label{eqelechole1}
\psi_e^T (\mathbf k) = \begin{pmatrix}
\psi_N^T (\mathbf k) & \frac{ \left( \varepsilon -\Omega \right) \, e^{-i\,\varphi }} {\Delta_0}\,
 \psi_N^T (\mathbf k)
\end{pmatrix} \text{ and }
\psi_h^T (\mathbf k) = \begin{pmatrix}
\psi_N^T (\mathbf k) & \frac{ \left( \varepsilon + \Omega \right) \, e^{-i\,\varphi }} {\Delta_0}\,
 \psi_N^T (\mathbf k)
\end{pmatrix} ,
\end{align}
respectively, where 
\begin{align}
\Omega = i \,\sqrt{ \Delta_0^2 - \varepsilon^2 } \,.
\end{align}
We also define the quantity
\begin{align}
\label{eqbeta}
\beta =\arccos(\varepsilon /\Delta_0)\,,
\end{align}
which will be used extensively in the expressions that follow.

Andreev reflection at a normal-metal–superconductor interface couples the electron-like and hole-like wavefunctions.
Here we follow the approach outlined in Ref.~\cite{titov-graphene}, where the quasiparticles and and the quasiholes are coupled locally by means of a boundary condition on the wavefunction in the normal region.
Now, the $x$-component of the probability-current density operator must be modified to
\begin{align}
\label{eqcur12}
\check{ \mathbf{j} }^{N}_x = 
\partial_{k_x} \mathcal{H}_{\text{BdG}}(\mathbf  k, \mathbf r)
= \mathcal{S}_x \otimes {\mathcal S}_0  
- {\mathcal S}_0 \otimes  \mathcal{S}_x \,.
\end{align} 
Therefore, for a wavefunction of the form $ \psi_{\text{BdG}, 0} \equiv  
\left( c_1  \quad 0  \quad c_3 \quad c_4 \quad 0 \quad c_6 \right ) ^T $, we have
$ \psi_{\text{BdG}, 0}^\dagger \, \check{ \mathbf{j} }^{N}_x \, \psi_{\text{BdG}, 0} = 0$. Analogous to the discussions of boundary conditions for purely normal-state junctions, we here infer that the electron- and hole-wavefunctions, corresponding to the flat-band state, drop out from the equations arising out of the boundary conditions. Furthermore, Eq.~\eqref{eqcur12} translates to the following constraints, useful for relating the components from a piecewise-defined wavefunction, $ \psi_{\text{BdG}} \equiv  
\left( c_1  \quad c_2  \quad c_3 \quad c_4 \quad c_5 \quad c_6 \right ) ^T $, across an interface at $x=x_0$:
\begin{align}
\label{eqbdy13}
& \lim_{ \delta \rightarrow 0 }
\left[c_1(x_0-\delta, y ) + c_3 (x_0-\delta, y ) \right ] = \lim_{ \delta \rightarrow 0 }
 \left [ c_1(x_0 + \delta, y ) + c_3 (x_0 +\delta, y ) \right ],
\lim_{ \delta \rightarrow 0 }  c_2(x_0-\delta, y ) 
= \lim_{ \delta \rightarrow 0 } c_2 (x_0 + \delta, y ) \,.
\nn & \lim_{ \delta \rightarrow 0 }
\left[c_4 (x_0-\delta ) + c_6 (x_0-\delta, y ) \right ] 
= \lim_{ \delta \rightarrow 0 }
 \left [ c_4 (x_0 + \delta, y ) + c_6 (x_0 +\delta, y ) \right ],
\lim_{ \delta \rightarrow 0 }  c_5 (x_0-\delta, y ) = 
\lim_{ \delta \rightarrow 0 } c_5 (x_0 + \delta, y ) \,.
\end{align}
Here, we have two boundaries: at $x_0 = 0$ and $x_0 = L$.

Because the propagation direction is along the $x$-axis, the translation symmetry is broken in that direction, whereas the transverse momentum component $k_y$ is conserved across the S-B and B-S junctions.
We denote the polar angle as $\theta =\arctan(k_y /k_x )$.
Using Eqs.~\eqref{eqwv1} and \eqref{eqelechole1}, let us now elucidate the form of the eigenfunction
$$ \Psi (\mathbf r, k_\perp) = 
\psi_{I} (\mathbf r, k_\perp) \,\Theta(-x)
 + \psi_{II} (\mathbf r , k_\perp) \,\Theta(x)\, \Theta(L-x) 
 +  \psi_{III} (\mathbf r , k_\perp) \,\Theta(x-L) \,,$$
expressed in a piecewise manner for the three regions, where we set the Fermi energy at $E_F$ for the corresponding normal states (i.e., for $\Delta_0 = 0 $) in the regions I and III.

\begin{enumerate}

\item In the right superconductor region, the wavefunction localizing at the
interface is described by a linear combination of the following form (as explained, for example, in chapter 5 of Ref.~\cite{asano}):
\begin{align}
\psi_{III} ( \mathbf r, k_\perp ) = a_{r}\,\psi_{er}  ( \mathbf r, \theta_r )
+ b_{r}\,\psi_{hr}  ( \mathbf r, \tilde \theta_r ) \,,
\end{align}
where
\begin{align}
 \psi_{er}   ( \mathbf r, \theta_r ) &= \frac{ 
e^{ i \left \lbrace    k_y\, y \, 
+ \, k_x^{er}\, (x-L) \right \rbrace  } }
{\sqrt 2}
\begin{pmatrix}
 e^{i \,\beta -i\, \theta_r}
&  \sqrt 2 \, e^{i \,\beta }   
 &  e^{i \,\beta + i \, \theta_r}
 & e^{-i\,   \varphi_2 -i\, \theta_r}
&  \sqrt 2 \, e^{-i\,   \varphi_2}  
&   e^{-i\,   \varphi_2 + i \, \theta_r}
\end{pmatrix} ^T ,\nn
\sin  \theta_r  \simeq  \frac{k_\perp } {E_F} & \,, \quad
k_x^{er } \simeq  k_{\rm{mod} }
+ i \,\kappa \,,\quad
k_{\rm{mod} } \simeq 
\sqrt{ E_F^2  - k_\perp^2 } \,,
\quad 
\kappa = \frac{E_F \,\Delta_0 \sin \beta } 
{ k_{\rm{mod} } } \,,
\quad
\tan \theta_r  \simeq \frac{k_\perp} { k_{\rm{mod} }}\,,
\end{align}
and
\begin{align}
 \psi_{hr}   ( \mathbf r, \tilde \theta_r ) & = \frac{ 
e^{ i \left \lbrace   k_y\, y \, + \, k_x^{hr}\, (x-L) \right \rbrace  } }
{\sqrt 2}
\begin{pmatrix}
 e^{ -i \,\beta -i\, \theta_r}
&  \sqrt 2 \, e^{ - i \,\beta }   
 &  e^{- i \,\beta + i \, \theta_r}
 & e^{-i\,   \varphi_2 -i\, \theta_r}
&  \sqrt 2 \, e^{-i\,   \varphi_2}  
&   e^{-i\,   \varphi_2 + i \, \theta_r}
\end{pmatrix}^T ,\nn
\sin  \tilde \theta_r & \simeq  \frac{k_\perp } {E_F} \,, \quad
k_x^{hr} \simeq  -\, k_{\rm{mod} }
+ i \,\kappa \,,
\quad
\tan \tilde \theta_r  \simeq \frac{k_\perp} { - \, k_{\rm{mod} }} \,.
\end{align}

\item

In the normal state region, the wavefunction comprises a linear combination with the parametrization
\begin{align}
\psi_{II}   ( \mathbf r, k_\perp ) & = a\,\psi_{e+}   ( \mathbf r, \theta_{n} )  
+ b\,\psi_{e-}  ( \mathbf r , \theta_{n} )
+ c\,\psi_{h+}  ( \mathbf r , \tilde \theta_{n} ) 
+ d\,\psi_{h-}  ( \mathbf r, \tilde \theta_{n} ) \,,
\end{align}
where
\begin{align}
\psi_{e+}  ( \mathbf r , \theta_{n}) &= 
e^{i \left(   k_y\, y \, + \, k_x^{e}\, x \right) } \, 
f_1(\theta_{n})  \,  , \quad
f_1(\theta_{n})  = 
\begin{pmatrix}
e^{-i\,\theta_n} 
& \sqrt 2  & e^{i\,\theta_n} 
& 0 & 0 & 0 
\end{pmatrix}^T , \quad
\psi_{e-}    ( \mathbf r , \theta_n)
=  e^{i \left(  k_y\, y \, -\, k_x^{e}\, x \right) } \, 
f_1  ( \pi-\theta_n)\,,  
\nn k_x^{e} & 
= -\, \sqrt{ \left(  V_0 -E_F -\varepsilon  \right )^2  - k_\perp^2 }, 
\quad \cos \theta_n =\frac{ k_x^{e}} 
{\varepsilon+E_F -V_0} \,,\quad
\sin  \theta_n =
 \frac{ k_\perp} {\varepsilon+E_F -V_0}\,,
\end{align}
and
\begin{align}
& \psi_{h+}   ( \mathbf r, \tilde  \theta_n  )   = 
e^{i \left(   k_y\, y \, +\, k_x^h\, x \right) } \,
f_2 (\tilde \theta_n) \,,\quad
f_2 (\tilde \theta_n) =
\begin{pmatrix}
 0 & 0 & 0 & e^{-i\,\tilde \theta_n}  & - \sqrt 2  & e^{i\, \tilde \theta_n}
\end{pmatrix}^T  ,\nn
& \psi_{h-}   ( \mathbf r, \tilde  \theta_n )   = 
e^{i \left(  k_y\, y \, -\, k_x^{h }\, x \right) } \, 
f_2 ( \pi - \tilde \theta_n)\,,
\quad k_x^h  
= \sqrt{ \left(  V_0 -E_F + \varepsilon  \right )^2  - k_\perp^2 }\,, 
\nn &  \cos \tilde \theta_n =\frac{ k_x^{h}} 
{\varepsilon-E_F +V_0} \,,\quad
\sin  \tilde \theta_n =
 \frac{ k_\perp} {\varepsilon - E_F + V_0 }\,.
\end{align}

\item In the left superconductor region, the linear combination, describing the wavefunction localizing at the
interface, must be expressed as
\begin{align}
\psi_{I}  ( \mathbf r, k_\perp ) = a_{l}\,\psi_{el}  ( \mathbf r, \theta_r )
+ b_{l}\,\psi_{hl}  ( \mathbf r, \tilde \theta_r ) \,,
\end{align}
where
\begin{align}
  \left \lbrace \psi_{el} ( \mathbf r, \theta_r ),  \, 
\psi_{hl} ( \mathbf r, \tilde \theta_r ) \right \rbrace 
= \left \lbrace \psi_{er} ( \mathbf r, \pi- \theta_r ), \,
 \psi_{hr} ( \mathbf r, \pi-\tilde  \theta_r ) 
 \right \rbrace \Big \vert_{\varphi_2 \rightarrow \varphi_1, \, (x-L) \rightarrow x}\,.
\end{align} 
This essentially amounts to flipping the signs of $\left \lbrace k_x^{er },  \,k_x^{hr } \right \rbrace $, because the left-moving electron-like and hole-like wavefunctions are of relevance in region I, which have the exponentially decaying behaviour \cite{asano}.

\end{enumerate}

Since the final results depend on the phase difference $\varphi_{12} \equiv \varphi_2 -\varphi_1 $, for simplification of the notations, we can set $\varphi_1 = 0$ and $\varphi_2 = \varphi_{12} $, without any loss of generality.
Imposing the continuity of the probability current density at the junctions located at $x = 0$ and $ x = L$, we get the following conditions [cf. Eq.~\eqref{eqbdy13}]:
\begin{align}
\label{eqbdy0}
& \left[ \psi_{I}  (0, y,  k_\perp) \right]_{1,1}
+ \left[ \psi_{I}  (0, y,  k_\perp) \right]_{1,3} = 
\left[ \psi_{II}  (0, y,  k_\perp) \right ]_{1,1}
+ \left[ \psi_{II}  (0, y,  k_\perp) \right ]_{1,3} ,
\quad \left[ \psi_{I}  (0, y,  k_\perp) \right]_{1,2} = 
\left[ \psi_{II}  (0, y,  k_\perp) \right ]_{1,2} ,\nn
& \left[ \psi_{I}  (0, y,  k_\perp) \right]_{1,4} + \left[ \psi_{I}  (0, y,  k_\perp) \right]_{1,6} = 
\left[ \psi_{II}  (0, y, k_\perp) \right ]_{1,4} + \left[ \psi_{II}  (0, y, k_\perp) \right ]_{1,6} ,
\quad \left[ \psi_{I}  (0, y,  k_\perp) \right]_{1,5} = 
\left[ \psi_{II}  (0, y,  k_\perp) \right ]_{1,5}, \nn
& \left[ \psi_{II}  ( L, y, k_\perp) \right]_{1,1} + \left[ \psi_{II}  ( L, y, k_\perp) \right]_{1,3} = 
\left[ \psi_{III}  (L,y, k_\perp) \right ]_{1,1} +\left[ \psi_{III}  (L,y, k_\perp) \right ]_{1,3} ,
\quad \left[ \psi_{II}  (L, y, k_\perp) \right]_{1,2} = 
\left[ \psi_{III}  (L, y, k_\perp) \right ]_{1,2} ,\nn
& \left[ \psi_{II}  (L, y, k_\perp) \right]_{1,4} + \left[ \psi_{II}  (L, y, k_\perp) \right]_{1,6} = 
\left[ \psi_{III}  (L, y, k_\perp) \right ]_{1,4} +\left[ \psi_{III}  (L, y, k_\perp) \right ]_{1,6} ,
\quad \left[ \psi_{II}  (L, y, k_\perp) \right]_{1,5} = 
\left[ \psi_{III}  (L, y, L, k_\perp) \right ]_{1,5}. \nn
\end{align}
We note that the subgap ABSs are localized near the S-B and B-S junctions with the localization length $ \sim \kappa^{-1}$, because they decay exponentially as we move away from the junction location into the bulk of one of the superconducting regions.

\subsection{Results}

The thin-barrier limit is physically equivalent to a Dirac-delta potential barrier. However, since we do not have any constraint on the derivatives of the componets of the wavefunction across the junctions, which follows from the linear-in-momentum dispersion \cite{urban, fang, ips3by2, krish-spin1}, the standard delta-function-potential approximation \cite{zagoskin,kwon_krish, ips-abs-semid} for thin barriers cannot be used here \cite{krish-moitri, ips_jj_rsw}. Instead, we need to start with Eq.~\eqref{eqbdy0}, and impose the approximations
\begin{align}
k_{x}^{e} \, L \rightarrow -\, \chi \text{ and  }
k_{x}^{h} \, L \rightarrow  \chi
\end{align}
in the exponential factors representing plane waves propagating along the $x$-axis.
Furthermore, the $\varepsilon $-dependence disappears from the polar angles, since $ -\theta_{n} \simeq \tilde \theta_{n} \simeq \arcsin \left(  \frac{V_0-E_F} { k_\perp}\right) $.

\begin{figure}[t]
\centering
\subfigure[]{\includegraphics[width=0.4 \textwidth]{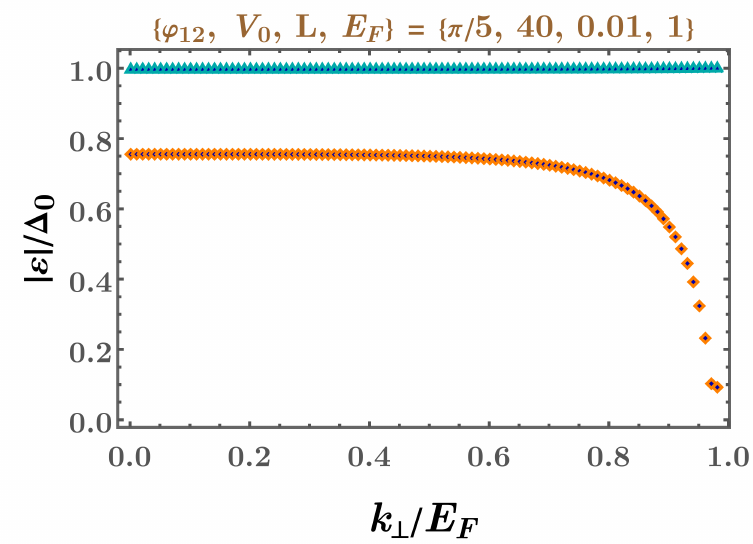}}\qquad
\subfigure[]{\includegraphics[width=0.4 \textwidth]{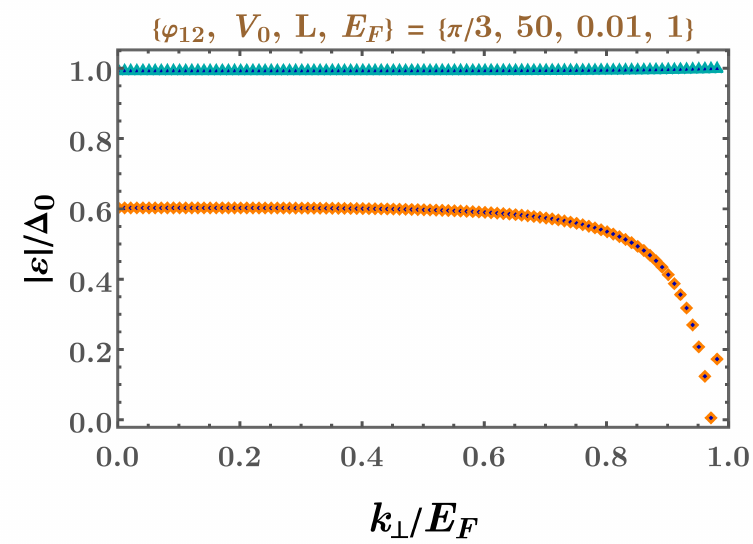}}
\subfigure[]{\includegraphics[width=0.4 \textwidth]{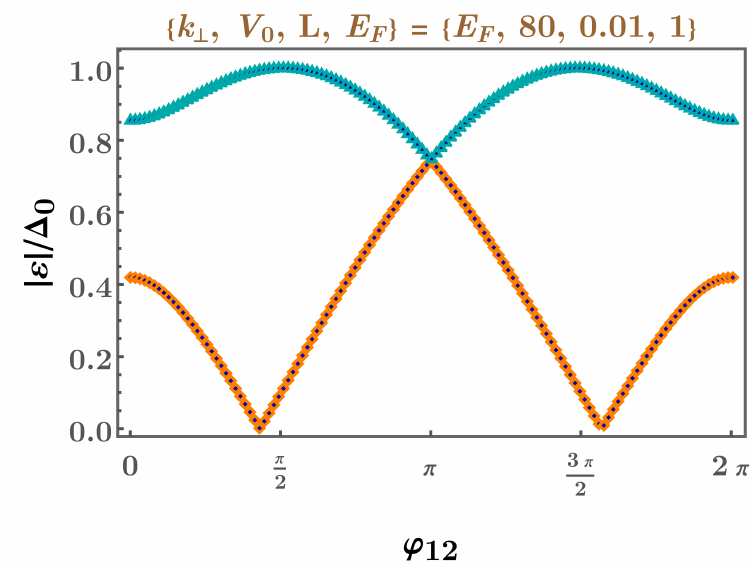}}\qquad
\subfigure[]{\includegraphics[width=0.4 \textwidth]{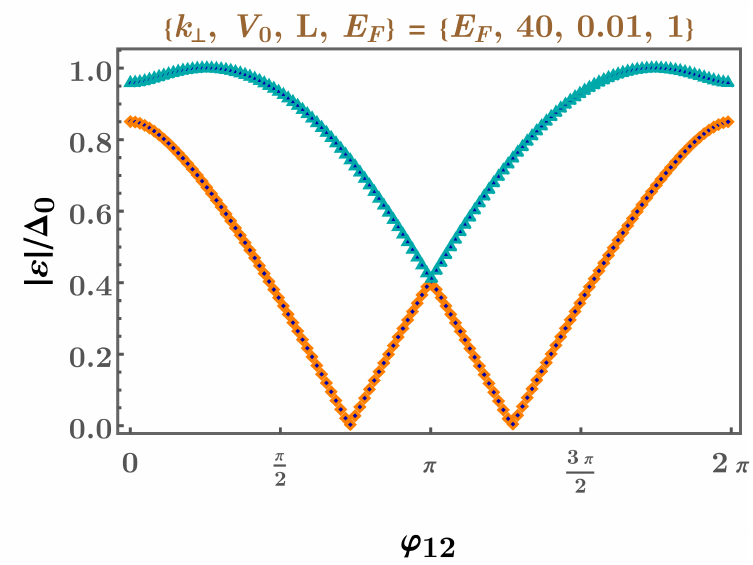}}
\caption{
Behaviour of $ |\varepsilon| $ against $k_\perp $ [subfigures (a) and (b)] and $\varphi_{12}$ [subfigures (c) and (d)], for some representative values of the remaining parameters (shown in the plotlabels).
\label{fige}}
\end{figure}

\begin{figure}[t]
\centering
\subfigure[]{\includegraphics[width=0.4 \textwidth]{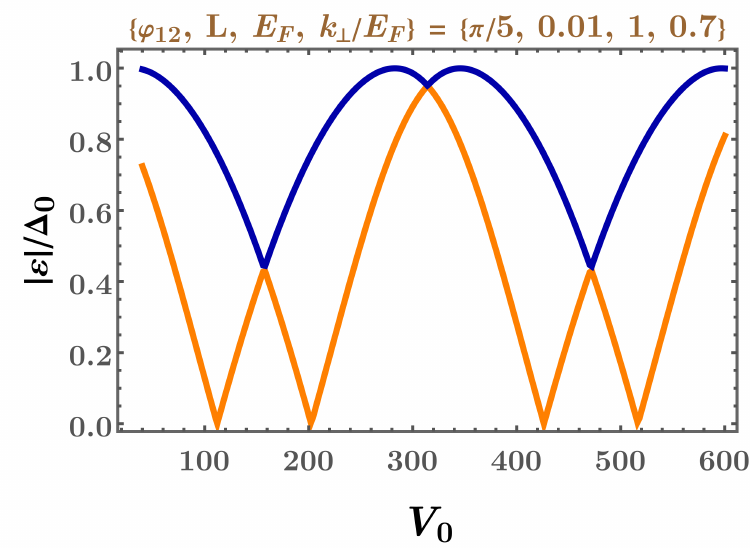}}\qquad
\subfigure[]{\includegraphics[width=0.4 \textwidth]{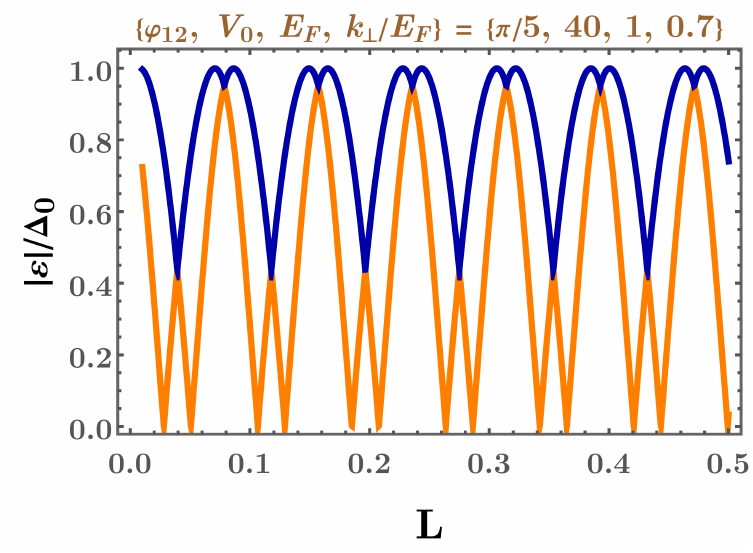}}
\caption{
Behaviour of $ |\varepsilon| $ against $ V_0 $ [subfigure (a)] and $L$ [subfigure (b)], for some representative values of the remaining parameters (shown in the plotlabels).
\label{figvl}}
\end{figure}

From the four combinations of the components of the wavefunction appearing at each of the two boundaries, as shown in Eq.~\eqref{eqbdy0}, we get
$2\times 4 = 8 $ linear homogeneous equations in the $ 8 $ variables $ \left \lbrace  a_{l}, \, b_{l},  \, a, \,b, \,
c, \, d, \, a_{r}, \, b_{r} \right \rbrace$, which constitute the $8$ unknown coefficients of the wavefunction. The overall $x$-independent factors of $e^{ i \, k_y\, y }$ cancel out equation by equation.
Let $M$ be the $ 8 \times 8 $ matrix constructed from the coefficients of these 8 variables, whose explicit form is given by
\begin{align}
& M \nn& = \begin{pmatrix} 
  - \, e^{i \, \beta} \cos \theta_r  &
   e^{-i \, \beta} \cos \theta _r  &  0 &  0 & 
 -\, \sqrt 2 \cos \theta _n  & \sqrt 2 \cos \theta _n  &  0 &  0  \\  
e^{i \, \beta} &  e^{-i \, \beta} &  0 &  0 & - \sqrt {2} & - \sqrt {2} &  0 &  0  \\
 -  \cos \theta _r  &  \cos \theta _r  &  0 &  0 &  0 &  0 &
 -  \, \sqrt {2} \cos \theta _n  &  \sqrt {2} \cos \theta _n   \\  
1 &  1 &  0 &  0 &  0 &  0  & \sqrt {2} &\sqrt {2}  \\
0 &  0 &  e^{i \, \beta} \cos \theta_r  
& - \, e^{-i \, \beta}  \cos \theta_r  & 
- \,\sqrt {2} \,  e^{i \, \chi}  \cos \theta _n  
&  \sqrt {2}  \, e^{-i \, \chi} \cos \theta _n  &  0 &  0  \\  
0 &  0 &   e^{i \, \beta} &  e^{-i \, \beta} 
& - \sqrt {2} \, e^{i \, \chi} & - \sqrt {2}  \, e^{-i \, \chi} &  0 &  0  \\  
0 &  0 &  e^{-i \, \varphi_{12}} \cos \theta_r & 
-  e^{-i \, \varphi_{12}}  \cos \theta_r  &  0 &  0 & 
-  \, \sqrt {2} \, e^{i \, \chi} \cos \theta_n  &  
 \sqrt {2}  \,e^{-i \, \chi}  \cos \theta_n   \\  
0 &  0 &  1 & 1 &  0 &  0 &
\sqrt {2}  \, e^{i \,\left(    \varphi_{12} + \chi \right )} 
& \sqrt {2} \, e^{i \, \left ( \varphi_{12} - \chi \right) }  
\end{pmatrix}.
\end{align}
The consistency of the 8 equations is ensured by the condition $\text{det} \,M = 0 $. After some row and column operations on the matrix $M$, we can bring it to the form of $\text{det}\, \tilde M = 0 $, where
\begin{align}
\label{eqmtilde}
& \tilde M \nn& = \begin{pmatrix} 
- \cos  \beta  \cos \theta_r & - i \sin  \beta \cos \theta_r  &  0 &  0 
&  \cos \theta _n &  0 &  0 &  0  \\  
i \sin  \beta & \cos  \beta &  0 &  0 &  0 &   1 &  0 &  0  \\ 
 - \cos \theta_r  &  0 &  0 &  0 &  0 &  0 &  -\cos \theta_n   &  0  \\
0 &  1 &  0 &  0 &  0  &  0 &  0 &  1  \\
0 &  0 & \cos  \beta \cos \theta_r  &  i \sin  \beta \cos \theta_r  &  \cos  \chi \cos \theta_n   
 &    i\sin  \chi \cos \theta_n  &  0 &  0  \\
 0 &  0 &  i\sin  \beta &\cos  \beta &   i \sin  \chi &  \cos  \chi &  0 &  0  \\
0 & 0 &  e^{-i\, \varphi_{12}} \cos \theta _r  &  0 &  0 &  0 & - \cos  \chi \cos \theta_n  &
  -   i\sin  \chi \cos \theta _n   \\
0 &  0 &  0 &  1 &  0 &  0 
&  i \, e^{i \,\varphi_ {12}} \sin \chi &  e^{i\,\varphi_ {12}} \cos \chi 
\end{pmatrix}.
\end{align}
\begin{figure}[t]
\centering
\subfigure[]{\includegraphics[width=0.4 \textwidth]{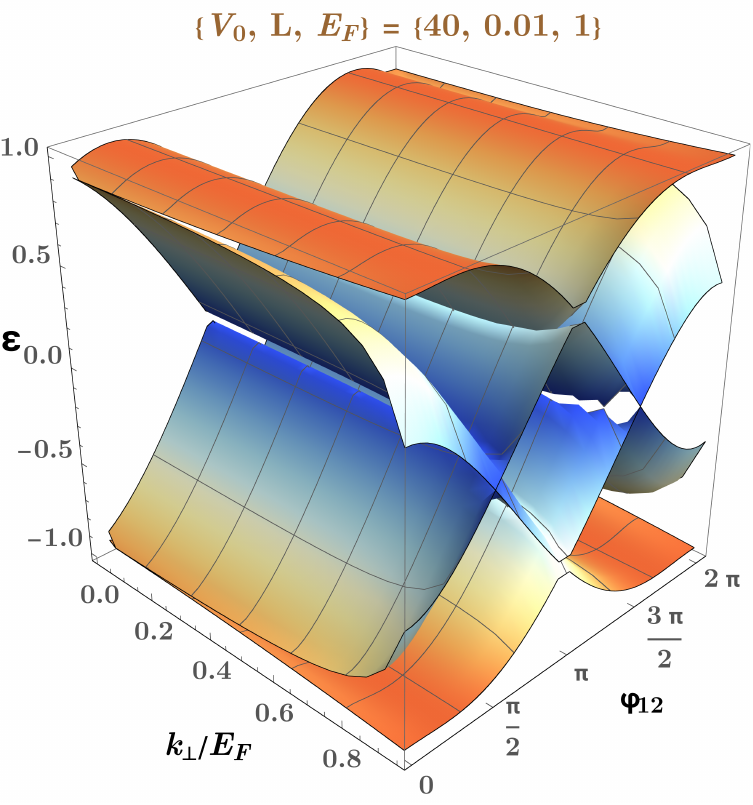}}\hspace{2 cm}
\subfigure[]{\includegraphics[width=0.4 \textwidth]{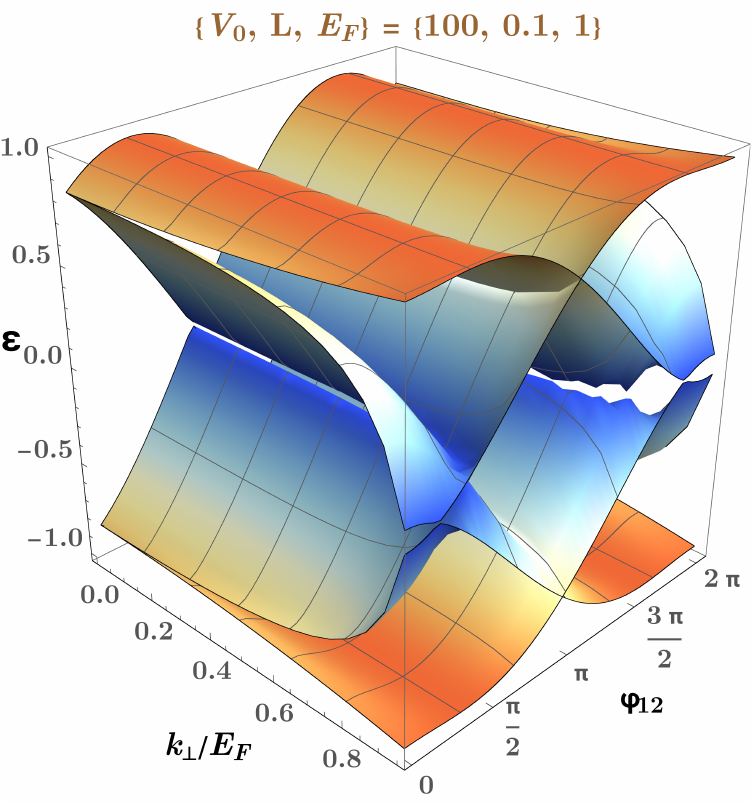}}
\subfigure[]{\includegraphics[width=0.4 \textwidth]{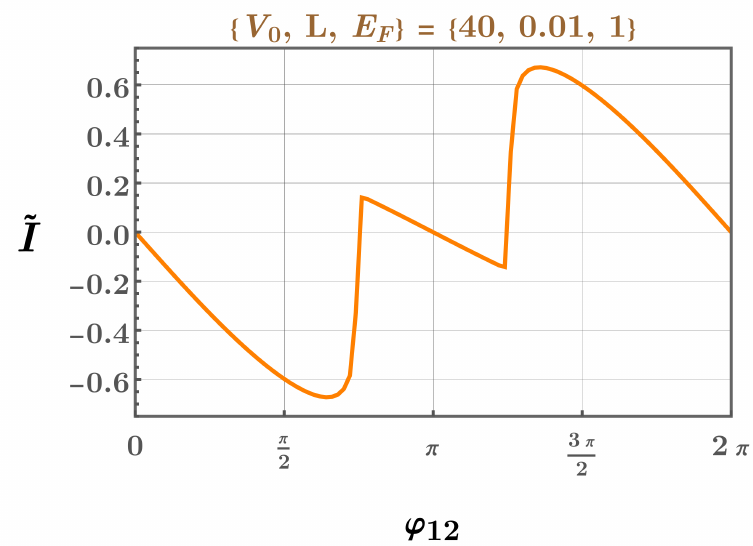}}\hspace{2 cm}
\subfigure[]{\includegraphics[width=0.4 \textwidth]{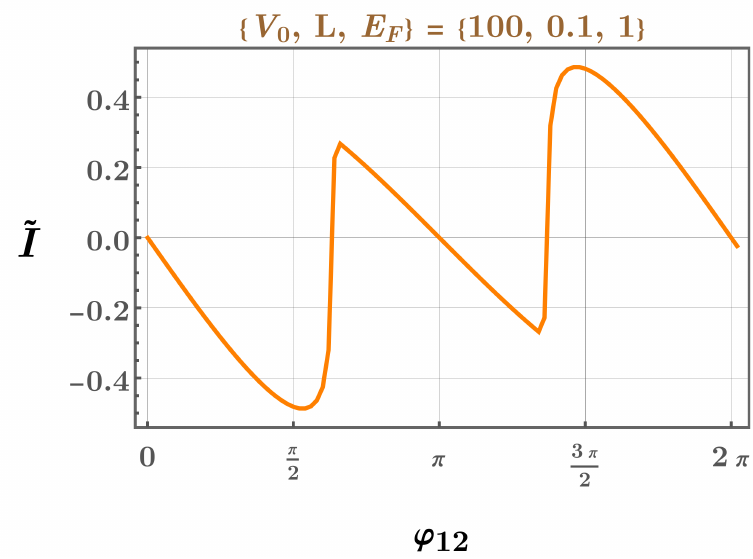}}
\caption{
The first two subfigures show the energy $\varepsilon $ of the two pairs of the Andreev bound states against the $k_\perp $-$\varphi_{12}$ plane, for (a) $V_0 = 50$, $L=0.01$, and $E_F = 1$; (b) $V_0 = 100$, $L=0.1$, and $E_F = 1$.
Subfigures (c) and (d) illustrate the behaviour of the total Josephson current ($\propto \tilde I $) in arbitrary units, as a function of $\varphi_{12}$, obtained at $k_B \, T = 0.005\, \Delta_0 $, for the same sets of parameter values as used in (a) and (b), respectively.
\label{figj}}
\end{figure}

The resulting equation takes the form of
\begin{align}
\label{eqabc0}
\mathcal A \sin (2\beta) +\mathcal{B} \cos (2\beta) + \mathcal C = 0\,,
\end{align}
where
\begin{align}
\label{eqabc}
& \mathcal A =  -2 \sin  (2\chi) \cos \theta_n   \cos \theta_r  
\left (\cos^2 \theta _n + \cos^2 \theta_r  \right), \nn &
\mathcal B =
\sin^2 \chi \left (\cos^4 \theta_n  + \cos^4 \theta_r   \right) 
- \left[ 3\cos  (2\chi) +  1 \right ] \cos^2 \theta_n \cos^2 \theta_r\,,
\nn & \mathcal C =
4 \cos \varphi  _ {12}  \cos^2 \theta_n   \cos^2 \theta_r  
- \sin^2 \chi \left (\cos^2 \theta_r   - \cos^2 \theta_n \right)^2\,.
\end{align}
Consequently, using Eq.~\eqref{eqbeta}, we get two pairs of ABS energy values equal to $\pm |\varepsilon|$, where
\begin{align}
\label{eqabs}
| \varepsilon | = \frac {\Delta_0 }  {\sqrt {2}}
\,\sqrt {1 - \frac { \mathcal B \,\mathcal  C  \, \pm \, 
| \mathcal  A | \, \sqrt {\mathcal A^2 + \mathcal  B^2 -\mathcal  C^2} } 
{\mathcal A^2 + \mathcal  B^2}}\,.
\end{align}
In Fig.~\ref{fige}, we show the variation of $|\varepsilon|$ as a function of (I) $ k_\perp $ (with $\varphi_{12} $ held fixed) and (II) $\varphi_{12} $ (with $  k_\perp $ held fixed), for some representative values of $V_0$ and $L$. Since $\mathcal C$ is a function of $\cos \varphi_{12}$ [cf. Eq.~\eqref{eqabc}], it follows from Eq.~\eqref{eqabs} that the subgap energies are periodic in
$\varphi_{12} $ with period $2\pi $, which is also evident from the curves shown in the lower panel of Fig.~\ref{fige}.
In Fig.~\ref{figvl}, we plot the dependence of $|\varepsilon|$ as a function of (I) $V_0$ (with $ L $ held fixed)
and (II) $ L $ (with $  V_0 $ held fixed), for some representative values of $k_\perp$ and $\varphi_{12}$. The curves show an oscillatory behaviour, with respect to both $V_0$ and $L$. This is not surprising, given that the solutions are functions of $\cos (2 \chi)$.
Lastly,, we have plotted the behaviour of $ \pm| \varepsilon | $ as functions of $\varphi_{12} $ and $ k_\perp$ in the upper panel of Fig.~\ref{figj}, thus illustrating the dependence of the subgap energies on both these variables in a combined way. Subfigures Fig.~\ref{figj}(a) and Fig.~\ref{figj}(b) show the behaviour for two sets of values of the remaining junction variables $V_0$ and $L$, with $E_F$ kept fixed at unity.

Restoring the $\hbar $ factors, the Josephson current density across the S-B-S junction, at a temperature $T$, is given by \cite{zagoskin,titov-graphene}
\begin{align}
I_J( \varphi_{12} ) = -\frac{2\, e}{\hbar}\,\frac{W} {2\,\pi} \sum_{n=1} ^ 4
\int  dk_y \, \frac{ \partial \epsilon_n}
{ \partial  \varphi_{12} } \,f(\epsilon_n) \,,
\end{align}
where $\epsilon_n$ labels the energy values of the four ABSs, arising from the two distinct values of $|\varepsilon|$. Here, $f(\zeta) = 1 / \left(  1+e^{\frac{\zeta}{k_B\, T}} \right)$ is the Fermi-Dirac distribution function, with $k_B$ being the Boltzmann constant. The lower panel of Fig.~\ref{figj} demonstrates the behaviour of $I_J$ as a function of $\varphi_{12} $, scaled by the appropriate numbers/variables (which we label as $\tilde I $), for two sets of representative parameters.

\section{3d model}
\label{sec3d}

The 3d pseudospin-1 quasiparticles are the generalizations of the pseudospin-1/2 quasiparticles in Weyl semimetals, with the nodal points acting as sources/sinks of the Berry flux, acting as Berry-curvature monopoles of magnitude 2 \cite{bernevig, spin13d1}. They can be realized in 3d tight-binding models for cold fermionic atoms in cubic optical lattices \cite{cold-atom} and, via ab initio simulations, have been identified in materials like TaN, NbN, and WC-type ZrTe \cite{spin13d1, spin13d2, spin13d3, spin13d4}. Analogous to Eq.~\eqref{eqham2d}, the effective low-energy continuum
Hamiltonian, in the vicinity of a nodal point, is given by
\begin{align}
\label{eqham3d}
\mathcal{H}_{3d}(\mathbf  k) = v_F \left(k_x \,\mathcal S_x + k_y \,\mathcal S_y +  k_z \,\mathcal S_z \right).
\end{align}
We again adopt the simplication of setting $v_F$ to unity.
The energy eigenvalues are again given by Eq.~\eqref{eqeigen}, with $\mathbf k$ now being equal to $ k_x \, \hat {\mathbf x} + k_y \, \hat {\mathbf y} + k_z \, \hat {\mathbf z}$.
A set of orthogonal eigenvectors is captured by
\begin{align}
\label{eqwv2}
\Psi_s (\mathbf k) & = \left(
\frac{2 \,k_z \left (k_z +s\, k \right )
+k_x^2 + k_y^2} {(k_x + i\, k_y)^2} \qquad 
\frac{\sqrt{2} \left (k_z + s\,k \right )} {k_x+ i\, k_y} \qquad 1
 \right )^T 
 \left(\text{for energy} = s\, k \right) \nn \text{ and }
  \Psi_0 (\mathbf k) & = \left ( \frac{-k_x + i\,  k_y}
 {k_x+ i\, k_y} \qquad \frac{\sqrt{2}\, k_z}{k_x+ i\, k_y}
 \qquad 1  \right )^T \left(\text{for the flat band}\right).
\end{align}

The 3d probability-current density operator takes the form of $ \check{ \mathbf{j} }_{3d} (\mathbf k)= \nabla_{\mathbf{k}} \mathcal{H}_{3d}(\mathbf  k)
= \mathcal{S}_x \, \hat {\mathbf x} + \mathcal{S}_y \, \hat {\mathbf y} + \mathcal{S}_z \, \hat {\mathbf z}$.
The current density vector from the wavefunction $\psi \equiv  \left( c_1  \quad c_2  \quad c_3 \right ) ^T $ is
\begin{align}
\label{eqcur2}
\psi^\dagger \, \check{ \mathbf{j} }_{3d} \,\psi =
\sqrt 2 \, {\text{Re} \left[ c_ 2^* \left ( c_ 1 + c_ 3 \right) \right]} 
\,  \hat {\mathbf x}
 + 
\sqrt 2 \, {\text{Im} \left[ c_ 2^* \left ( c_ 1 - c_ 3 \right) \right]} 
\,  \hat {\mathbf y}
+ \left[ \left |  c_ 1 \right |^2 - \left | c_ 3 \right |^2  \right]
\,  \hat {\mathbf z}\,.
\end{align}
This again leads to a vanishing contribution from the flat-band wavefunction.
Eq.~\eqref{eqcur2} also tells us that if the quasiparticles encounter a perpendicular barrier, say at $z = z_0$, while propagating along the $z$-direction, the conservation of the probability-current density leads to the two boundary conditions \cite{krish-spin1},
\begin{align}
\label{eqbdy2}
\lim_{ \delta \rightarrow 0 }  c_1 (x, y, z_0-\delta ) = \lim_{ \delta \rightarrow 0 } c_1 (x, y, z_0 + \delta ) 
\text{ and }
\lim_{ \delta \rightarrow 0 }  c_3 ( x, y, z_0-\delta ) = 
\lim_{ \delta \rightarrow 0 } c_3 ( x, y, z_0 + \delta )\,.
\end{align}
In this case, if the Fermi energy cuts the dispersion profile at an energy value $ E $, the wavefunction will have the exponential factor $e^{i\, \text{sgn}(E) \,k_z \, z}$, where $ k_z  = \sqrt{ E^2-k_\perp^2 }$ and $k_\perp = \sqrt{ k_x^2 + k_y^2 } $.

\subsection{S-B-S junction's interfaces oriented perpendicular to the $z$-direction}
\label{secsbs3d}

Here we consider the propagation of the quasiparticles/quasiholes along the $z$-axis. We would like to point out that the final results would be the same if we had taken the propagation direction to be along the $x$- or $y$-axis, which follows from the isotropic nature of the dispersion. In other words, we can freely rotate our coordinate system without any physically observable consequences. The reason for our particular orientation of the coordinate axes is that the expressions for the wavefunctions and the resulting intermediate equations (to be solved) turn out to be simpler for making this choice.

Setting up the S-B-S configuration as shown in Fig.~\ref{figsetup}(b), we model the superconducting pair potential as
\begin{align}
\label{eqscorder}
\Delta (z) =\begin{cases} 
\Delta_0\,e^{i\,\varphi_1 }\,{\mathcal S}_0 
 &\text{ for }  z\leq 0   \\
0 &\text{ for } 0 < z< L \\
\Delta_0\,e^{i\,\varphi_2 } \,{\mathcal S}_0  &\text{ for } z \geq L
\end{cases}, \quad
{\mathcal S}_0 = {\mathbb{1}}_{3\times 3} \,,
\end{align}
representing Cooper pairing in the s-wave channel. Due to the presence of the barrier region, we need to consider the potential energy
\begin{align}
V(z)
=\begin{cases} 
0
 &\text{ for }  z\leq 0 \text{ and } z\geq L  \\
 V_0 &\text{ for } 0 < z< L 
\end{cases}.
\end{align} 
The resulting BdG Hamiltonian is given by
\begin{align}
& H = \frac{1} {2} \sum_{\mathbf k} \Psi^\dagger_{\mathbf k} \,H_{\text{BdG}} (\mathbf k)
\Psi_{\mathbf k} , \quad
\Psi_{\mathbf k} = \begin{pmatrix}
c_1 (\mathbf k) & c_2 (\mathbf k) & c_3 (\mathbf k)  & 
c_1^\dagger (-\mathbf k) & c_2 ^\dagger(-\mathbf k) & c_3 ^\dagger(-\mathbf k) 
\end{pmatrix}^T\,, \nn
& H_{\text{BdG}} (\mathbf k) =
\begin{pmatrix}
\mathcal{H}_{3d}(\mathbf  k) -E_F + V(z)  & \Delta(z)  \\ 
 \Delta^\dagger(z) &  E_F- V(z) -\mathcal{H}_{3d}^T(-\mathbf  k)  \\  
\end{pmatrix},
\label{eq_bdg}
\end{align}
where the subscripts $\lbrace 1, 2, 3\rbrace $ on the fermionic creation and annihilation operators represent the three distinct band indices. Analogous to the 3d case, we assume the energy-scale hierarchies of $V_0 \gg E_F  \gg \Delta_0 $ and $(V_0-E_F) \gg E_F $.

The electron-like and the hole-like BdG quasiparticles are obtained from the eigenvalue equation
\begin{align}
H_{\text{BdG}} ( \mathbf k\rightarrow -i \boldsymbol{\nabla}_{\mathbf r}) \,\psi_{\mathbf k} (\mathbf r) 
= \varepsilon \, \psi_{\mathbf k} (\mathbf r) \,,
\end{align}
where $\mathbf r =  x \, \hat {\mathbf x} + y \, \hat {\mathbf y} +  z \, \hat {\mathbf z}$ is the 3d position vector. If $ \psi_N (\mathbf k)$ is an eigenfunction of $\mathcal{H}_{3d}(\mathbf  k)$ (with the superconducting phase factor of $\varphi $), then the electron-like and hole-like eigenfunctions of $ H_{\text{BdG}} ( \mathbf k) $ are given by the same relations as shown in Eq.~\eqref{eqelechole1}.

For considering normal-metal–superconductor interfaces, the $z$-component of the probability-current density operator must be modified to
\begin{align}
\label{eqcur22}
\check{ \mathbf{j} }^{N}_z = 
\partial_{k_z} \mathcal{H}_{\text{BdG}}(\mathbf  k, \mathbf r)
= \mathcal{S}_z \otimes {\mathcal S}_0  
- {\mathcal S}_0 \otimes  \mathcal{S}_z \,.
\end{align} 
Analogous to the 2d case, we here infer that the electron- and hole-wavefunctions, corresponding to the flat-band state, drop out from the equations arising out of the boundary conditions. In addition, Eq.~\eqref{eqcur22} translates into the following constraints, useful for relating the components from a piecewise-defined wavefunction, $ \psi_{\text{BdG}} \equiv  
\left( c_1  \quad c_2  \quad c_3 \quad c_4 \quad c_5 \quad c_6 \right ) ^T $, across an interface at $z=z_0$:
\begin{align}
\label{eqbdy3d}
& \lim_{ \delta \rightarrow 0 }  c_1 (x, y, z_0-\delta ) = 
\lim_{ \delta \rightarrow 0 } c_1 (x, y, z_0 + \delta ) \,, \quad
\lim_{ \delta \rightarrow 0 }  c_3 (x, y, z_0-\delta ) = 
\lim_{ \delta \rightarrow 0 } c_3 (x, y, z_0 + \delta )\,,\nonumber \\
& \lim_{ \delta \rightarrow 0 }  c_4 (x, y, z_0-\delta ) = 
\lim_{ \delta \rightarrow 0 } c_4 (x, y, z_0 + \delta ) \,, \quad
\lim_{ \delta \rightarrow 0 }  c_6 (x, y, z_0-\delta ) = 
\lim_{ \delta \rightarrow 0 } c_6 (x, y, z_0 + \delta )\,.
\end{align}
Here, we have the two boundaries at $ z_0 = 0$ and $ z_0 = L$.

Since the propagation direction is along the $z$-axis, the translation symmetry is broken in that direction, whereas the transverse momentum components $k_x$ and $k_y $ are conserved across the S-B and B-S junctions.
We denote the azimuthal angle in the $k_x k_y$-plane as $\phi =\arctan(k_y /k_x )$.
Using Eqs.~\eqref{eqwv2} and \eqref{eqelechole1}, let us now elucidate the form of the eigenfunction
$$ \Psi (\mathbf r, k_\perp) = 
\psi_{I} (\mathbf r, k_\perp) \,\Theta(-z)
 + \psi_{II} (\mathbf r , k_\perp) \,\Theta(z)\, \Theta(L-z) 
 +  \psi_{III} (\mathbf r , k_\perp) \,\Theta(z-L) \,,$$
expressed in a piecewise manner for the three regions, where we set the Fermi energy at $E_F$ for the corresponding normal states (i.e., for $\Delta_0 = 0 $) in the regions I and III.

\begin{enumerate}

\item In the right superconductor region, the wavefunction localizing at the
interface is described by a linear combination of the following form (see chapter 5 of Ref.~\cite{asano}):
\begin{align}
\psi_{III} ( \mathbf r, k_\perp ) = a_{r}\,\psi_{er}  ( \mathbf r, \theta_r )
+ b_{r}\,\psi_{hr}  ( \mathbf r, \tilde \theta_r ) \,,
\end{align}
where

\begin{align}
 \psi_{er}^T   ( \mathbf r, \theta_r ) &=  
e^{ i \left \lbrace  k_x\, x \, + \, k_y\, y \, 
+ \, k_{z}^{er}\, (z-L) \right \rbrace  } \,
e^{- i \,\phi \, {\mathcal S}_0 \otimes {\mathcal S}_z }
\nn &  \hspace{0.75 cm} \times
\begin{pmatrix}
\frac{ e^{i\,\beta }
\left( 1 + \cos \theta_r \right )} {\sqrt 2}
&  e^{i\,\beta } \sin \theta_r 
& \frac{ e^{i\,\beta }\left(  1 - \cos \theta_r \right)
} {\sqrt 2}
& \frac{ e^{-i\,   \varphi_2} \left( 1 + \cos \theta_r \right) 
} {\sqrt 2}
& e^{-i\,   \varphi_2}  \sin \theta_r 
& \frac{ e^{-i\,   \varphi_2} \left( 1 - \cos \theta_r  \right) 
} {\sqrt 2}
\end{pmatrix}  ,\nn
\sin  \theta_r  \simeq  \frac{k_\perp } {E_F} & \,, \quad
k_{z}^{er } \simeq  k_{\rm{mod} }
+ i \,\kappa \,,\quad
k_{\rm{mod} } \simeq 
\sqrt{ E_F^2  - k_\perp^2 } \,,
\quad 
\kappa = \frac{E_F \,\Delta_0 \sin \beta } 
{ k_{\rm{mod} } } \,,
\quad
\tan \theta_r  \simeq \frac{k_\perp} { k_{\rm{mod} }}\,,
\end{align}

\begin{align}
 \psi_{hr}^T   ( \mathbf r, \tilde \theta_r ) & = 
e^{ i \left \lbrace k_x\, x \, + \, k_y\, y \, + \, k_{z}^{hr}\, (z-L) \right \rbrace 
 } \,
e^{- i \,\phi \,{\mathcal S}_0 \otimes {\mathcal S}_z }
\nn &  \hspace{0.4 cm}  \times
\begin{pmatrix}
\frac{ e^{ - i\,\beta }
\left( 1 + \cos \tilde \theta_r \right ) 
} {\sqrt 2}
& e^{ -i\,\beta } \,\sin \tilde \theta_r 
& \frac{ e^{ - i\,\beta }\left(  1 - \cos \tilde \theta_r \right)
} {\sqrt 2}
& \frac{e^{-i\,   \varphi_2} \left( 1 + \cos  \tilde \theta_r \right) 
} {\sqrt 2}
& e^{-i\,   \varphi_2}  \sin \theta_r 
& \frac{ e^{-i\,   \varphi_2} \left( 1 - \cos \tilde \theta_r  \right) 
} {\sqrt 2}
\end{pmatrix} ,\nn
\sin  \tilde \theta_r & \simeq  \frac{k_\perp } {E_F} \,, \quad
k_{z}^{hr } \simeq  -\, k_{\rm{mod} }
+ i \,\kappa \,,
\quad
\tan \tilde \theta_r  \simeq \frac{k_\perp} { -\, k_{\rm{mod} }} \,,
\end{align}

\item

In the normal state region, we will have a linear combination of the following form:
\begin{align}
\psi_{II}   ( \mathbf r, k_\perp ) & = a\,\psi_{e+}   ( \mathbf r, \theta_{n} )  
+ b\,\psi_{e-}  ( \mathbf r , \theta_{n} )
+ c\,\psi_{h+}  ( \mathbf r , \tilde \theta_{n} ) 
+ d\,\psi_{h-}  ( \mathbf r, \tilde \theta_{n} ) \,,
\end{align}
where
\begin{align}
\psi_{e+}^T  ( \mathbf r , \theta_{n}) &= 
e^{i \left(  k_x\, x \, + \, k_y\, y \, + \, k_{z}^{e}\, z \right) } \, 
f_1(\theta_{n})   \,,\quad
f_1(\theta_{n})  = e^{- i \,\phi \,{\mathcal S}_0 \otimes {\mathcal S}_z }
\begin{pmatrix}
1 + \cos \theta_n & \sqrt {2}\, \sin \theta_n & 1 - \cos \theta_n 
& 0 & 0 & 0
\end{pmatrix} , \nn
\psi_{e-} ^T  ( \mathbf r , \theta_n)
&=  e^{i \left(  k_x\, x \, + \, k_y\, y \, -\, k_{z}^e\, z \right) } \, 
f_1  ( \pi-\theta_n)\,,  
\nn k_{z}^{e} & 
= -\, \sqrt{ \left(  V_0 -E_F -\varepsilon  \right )^2  - k_\perp^2 }, 
\quad \cos \theta_n =\frac{ k_{z}^{e}} 
{\varepsilon+E_F -V_0} \,,\quad
\sin  \theta_n =
 \frac{ k_\perp} {\varepsilon+E_F -V_0}\,,
\end{align}

\begin{align}
& \psi_{h+} ^T  ( \mathbf r, \tilde  \theta_n  )   = 
e^{i \left(  k_x\, x \, + \, k_y\, y \, +\, k_{z}^{h }\, z \right) } \,
f_2 (\tilde \theta_n) \,,\quad
f_2 (\tilde \theta_n) =
e^{- i \,\phi \, {\mathcal S}_0 \otimes {\mathcal S}_z }
\begin{pmatrix}
0 & 0 & 0 &
1 + \cos \tilde \theta_n & -\sqrt {2}\, \sin \tilde \theta_n & 1 - \cos \tilde \theta_n 
\end{pmatrix},\nn
& \psi_{h-} ^T  ( \mathbf r, \tilde  \theta_n )   = 
e^{i \left(  k_x\, x \, + \, k_y\, y \, -\, k_{z}^{h }\, z \right) } \, 
f_2 ( \pi - \tilde \theta_n)\,,
\nn & k_{z}^h  
= \sqrt{ \left(  V_0 -E_F + \varepsilon  \right )^2  - k_\perp^2 }, 
\quad  \cos \tilde \theta_n =\frac{ k_{z}^{h}} 
{\varepsilon-E_F +V_0} \,,\quad
\sin  \tilde \theta_n =
 \frac{ k_\perp} {\varepsilon - E_F + V_0 }\,,
\end{align}

\item In the left superconductor region, we will have a linear combination of the following form:
\begin{align}
\psi_{I}  ( \mathbf r, k_\perp ) = a_{l}\,\psi_{el}  ( \mathbf r, \theta_r )
+ b_{l}\,\psi_{hl}  ( \mathbf r, \tilde \theta_r ) \,,
\end{align}
where

\begin{align}
  \left \lbrace \psi_{el} ( \mathbf r, \theta_r ),  \, 
\psi_{hl} ( \mathbf r, \tilde \theta_r ) \right \rbrace 
= \left \lbrace \psi_{er} ( \mathbf r, \pi- \theta_r ), \,
 \psi_{hr} ( \mathbf r, \pi-\tilde  \theta_r ) 
 \right \rbrace \Big \vert_{\varphi_2 \rightarrow \varphi_1, \, (z-L) \rightarrow z}\,.
\end{align} 
This amounts to flipping the signs of $\left \lbrace k_{z}^{er },  \,k_{z}^{hr } \right \rbrace $, which is because we need to consider here the left-moving electron-like and hole-like wavefunctions \cite{asano}. The ``left-moving'' wavefunctions are physically admissible in this region, because they are the ones which decay exponentially.

\end{enumerate}

Since the final results depend on the phase difference $\varphi_{12} \equiv \varphi_2 -\varphi_1 $, for simplification of the notations, we set $\varphi_1 = 0$ and $\varphi_2 = \varphi_{12} $. Imposing the continuity of the probability current density at the junctions located at $z=0$ and $z=L$, we get the following conditions [cf. Eq.~\eqref{eqbdy3d}]:
\begin{align}
\label{eqbdy}
& \left[ \psi_{I}  (x, y, 0, k_\perp) \right]_{1,1} = 
\left[ \psi_{II}  (x, y, 0, k_\perp) \right ]_{1,1} ,
\quad \left[ \psi_{I}  (x, y, 0, k_\perp) \right]_{1,3} = 
\left[ \psi_{II}  (x, y, 0, k_\perp) \right ]_{1,3} ,\nn
& \left[ \psi_{I}  (x, y, 0, k_\perp) \right]_{1,4} = 
\left[ \psi_{II}  (x, y, 0, k_\perp) \right ]_{1,4} ,
\quad \left[ \psi_{I}  (x, y, 0, k_\perp) \right]_{1,6} = 
\left[ \psi_{II}  (x, y, 0, k_\perp) \right ]_{1,6}, \nn
& \left[ \psi_{II}  (x, y, L, k_\perp) \right]_{1,1} = 
\left[ \psi_{III}  (x, y, L, k_\perp) \right ]_{1,1} ,
\quad \left[ \psi_{II}  (x, y, L, k_\perp) \right]_{1,3} = 
\left[ \psi_{III}  (x, y, L, k_\perp) \right ]_{1,3} ,\nn
& \left[ \psi_{II}  (x, y, L, k_\perp) \right]_{1,4} = 
\left[ \psi_{III}  (x, y, L, k_\perp) \right ]_{1,4} ,
\quad \left[ \psi_{II}  (x, y, L, k_\perp) \right]_{1,6} = 
\left[ \psi_{III}  (x, y, L, k_\perp) \right ]_{1,6}.
\end{align}

\subsection{Results}

In order to implement the thin-barrier limit, we start with Eq.~\eqref{eqbdy}, and impose the approximations
\begin{align}
k_{z}^{e} \, L \rightarrow -\, \chi \text{ and  }
k_{z}^{h} \, L \rightarrow  \chi
\end{align}
in the exponential factors representing plane waves propagating along the $z$-axis, as explained in the 2d case.
Here too, the $\varepsilon $-dependence disappears from the polar angles, since $ -\theta_{n} \simeq \tilde \theta_{n} \simeq \arcsin \left(  \frac{V_0-E_F} { k_\perp}\right) $.

From the four combinations of the components of the wavefunction appearing at each of the two boundaries, as shown in Eq.~\eqref{eqbdy}, we get
$2\times 4 = 8 $ linear homogeneous equations in the $ 8 $ variables $ \left \lbrace  a_{l}, \, b_{l},  \, a, \,b, \,
c, \, d, \, a_{r}, \, b_{r} \right \rbrace$, which constitute the $8$ unknown coefficients of the wavefunction. In the resulting equations, both the overall $z$-independent factors of $e^{ i \left( k_x\, x \, + \, k_y\, y \right)}$, and the phase factors introduced by the action of $e^{- i \,\phi \, {\mathcal{S}}_0 \otimes {\mathcal{S}}_z}$, cancel out component by component.
Let $ \check{M}$ be the $ 8 \times 8 $ matrix constructed from the coefficients of these 8 variables, whose explicit form is given by
\begin{align}
\label{eqm3d}
& \check{M} \nn& = \begin{pmatrix} 
\frac { -e^{i \, \beta} \sin^2\big (\frac {\theta_r } {2} \big)} {\sqrt {2}} 
&\frac {-e^{-i\,\beta}\cos^2\big (\frac {\theta_r } {2} \big)} {\sqrt {2}} &  0 &  0 
&  \cos^2\big (\frac {\theta_n } {2} \big) &  \sin^2\big (\frac {\theta_n } {2} \big) &  0 &  0  \\
\frac { - e^{i \, \beta} \cos^2\big (\frac {\theta_r } {2} \big)} {2 \, \sqrt {2}} 
&\frac { - e^{-i\beta}\sin^2\big (\frac {\theta_r } {2} \big)} { 2 \, \sqrt {2} } &  0 &  0 
&  \frac {\sin^2\big (\frac {\theta_n } {2} \big) } {2}
&  \frac {\cos^2\big (\frac {\theta_n } {2} \big)} {2} &  0 &  0  \\
\frac { - \sin^2 \big (\frac {\theta_r } {2} \big)} {\sqrt {2}} 
&\frac { - \cos^2\big (\frac {\theta_r } {2} \big)} {\sqrt {2}} &  0 &  0
 &  0 &  0 &  \sin^2\big (\frac {\theta_n } {2} \big) &  \cos^2\big (\frac {\theta_n } {2} \big)  \\
\frac { - \cos^2\big (\frac {\theta_r } {2} \big)} {\sqrt {2}} 
&\frac { - \sin^2\big (\frac {\theta_r } {2} \big)} {\sqrt {2}} &  0 &  0 &  
0 &  0 &  \cos^2\big (\frac {\theta_n } {2} \big) 
&  \sin^2\big (\frac {\theta_n } {2} \big)  \\  
0 &  0 & \frac { - e^{i \, \beta} \cos^2\big (\frac {\theta_r } {2} \big)} {\sqrt {2}} 
&\frac { - e^{-i\, \beta}\sin^2\big (\frac {\theta_r } {2} \big)} {\sqrt {2}} 
&  e^{ i \, \chi }\cos^2\big (\frac {\theta_n } {2} \big) 
&  e^{- i \, \chi }\sin^2\big (\frac {\theta_n } {2} \big) &  0 &  0  \\ 
0 &  0 &\frac { - e^{i \, \beta} \sin^2\big (\frac {\theta_r } {2} \big)} {\sqrt {2}} 
&\frac { -e^{-i\,\beta}\cos^2\big (\frac {\theta_r } {2} \big)} {\sqrt {2}} 
&   e^{ i \, \chi }\sin^2\big (\frac {\theta_n } {2} \big) 
&   e^{- i \, \chi }\cos^2\big (\frac {\theta_n } {2} \big) &  0 &  0  \\
 0 &  0 & \frac { -  e^{i \,\varphi_{12} } \cos^2\big (\frac {\theta_r } {2} \big)} {\sqrt {2}} 
&\frac { -  e^{i \,\varphi_{12} }\sin^2\big (\frac {\theta_r } {2} \big)} {\sqrt {2}} &  0 &  0
 & e^{i  \,   \chi  } \sin^2\big (\frac {\theta_n } {2} \big)
 &  e^{ -i \, \chi } \cos^2\big (\frac {\theta_n } {2} \big)  \\  
  0 &  0 & \frac { - e^{i \,\varphi_{12} }
  \sin^2\big (\frac {\theta_r } {2} \big)} {\sqrt {2}} 
& \frac { -  e^{i \,\varphi_{12} } \cos^2\big (\frac {\theta_r } {2} \big)} {\sqrt {2}} &  0 &  0 & 
 e^{i  \,   \chi  }\cos^2\big (\frac {\theta_n } {2} \big) 
&  e^{ - i  \chi  }\sin^2\big (\frac {\theta_n } {2} \big)  
\end{pmatrix}.
\end{align}
The consistency of the 8 equations is ensured by the condition $\text{det} \,\check{M} = 0 $. After some row and column operations on the matrix $\check M$, we can bring the constraint equation to a simple form which turns out to be exactly the same as Eq.~\eqref{eqabc0}. Hence, the results for the 2d case ABSs are valid for the 3d case as well. This is similar to the scenario for the pseudospin-1/2 quasiparticles/quasiholes for graphene and Weyl semimetal, where the ABS-energy expressions are identical (as we can see from Eq.~(20) of Ref.~\cite{krish-moitri} and Eq.~(13) of Ref.~\cite{debabrata-krish}). Such outcomes are not surprising because, after all, both the 2d and 3d versions harbour excitations carrying the same pseudospin-values, with the continuum-limit dispersion having the same isotropic linear-in-momentum character.

The only difference from the 2d case is that, because the transverse part of the semimetallic slab is now two-dimensional (with a cross-section area of $W^2$), the expression for the Josephson current density across the weak link is now given by \cite{zagoskin,titov-graphene}
\begin{align}
\check I_J(\varphi_{12}) = -\frac{2\, e}{\hbar}\,\frac{W^2} {(2\,\pi)^2} \sum_{n=1} ^ 4
\int  dk_x\,dk_y \, \frac{ \partial \epsilon_n}
{ \partial \varphi_{12}} \,f(\epsilon_n) \,,
\end{align}
where the integration is now over a 2d momentum space (analogous to the cases of Refs.~\cite{debabrata-krish, ips_jj_rsw}).

\section{Summary and outlook}
\label{secsum}

To summarize, we have explicitly computed the expressions for energy values of the subgap ABSs in Josephson junctions, built with 2d and 3d triple-point fermions, in the thin-barrier limit. We have assumed a weak and homogeneous s-wave pairing in each superconducting region, which can be created via proximity effect by placing a superconducting electrode near it \cite{proximity-sc}. The barrier region can be implemented by applying a voltage of magnitude $V_0 $ across a piece of semimetal in its normal state. By constructing the appropriate BdG Hamiltonian, we have determined the wavefunction in a piecewise continuous manner, which localizes near the S-B and B-S junctions. On enforcing the consistency of the equations obtained from matching the boundary conditions, which results from the vanishing of the relevant determinant, we have found an equation consisting of the trigonometric functions $\cos (2\beta)$ and $\sin (2\beta )$. In fact, this equation is the same for both the 2d and 3d cases. The solutions for $\cos \beta$ give the absolute energy values $ |\varepsilon| $ of the ABSs. We have shown the closed-form analytical expressions for $|\varepsilon|$, which turn out to be two distinct functions [cf. Eq.~\eqref{eqabs}].

We would like to point out that, although each of the two $| \varepsilon |$-expressions is a function of $\cos \varphi_{12}$, the dependence is \textit{not} of the form $ \epsilon_{abs}^{lin} \equiv \Delta_0 \, \sqrt{1 - T_N\, \sin^2 \left( \varphi_{12} / 2\right) }$, found for two-band semimetals when the propagation direction is along a liearly dispersing direction \cite{titov-graphene,krish-moitri,debabrata-krish} (as discussed in the introduction section). This follows from the fact that increasing the number of bands changes the equations to be solved in the variables $\sin (2\beta)$ and $\cos(2\beta)$, arising as trigonomatric functions  of $\beta$. More explicitly, when the equation reduces to the simple form of $ \tilde{\mathcal{B} } \cos(2\beta) + \tilde{\mathcal{ C }} \cos \varphi_{12} =0 $, solutions of the form  $ \epsilon_{abs}^{lin} $ emerge. However, for multifold fermions, such a simplicity is lost, giving rise to multiple solutions of $|\varepsilon|$ as another artifact (see, for example Ref.~\cite{ips_jj_rsw}, where the case of an isotropic RSW semimetal with fourfold-degenerate nodal points has been studied). An analogous increase in the complexity of the equation is also caused when the propagation direction is along a momentum direction, along which the semimetal dispersion possesses a nonlinear dependence (as seen in Ref.~\cite{ips-abs-semid}, where the case of a semi-Dirac semimetal has been investigated).

If the barrier has impurities, structural defects, or non-uniformity, it will warrant incorporating that in the computations, which can be realistically achieved by resorting to numerics. In some simple cases, e.g., a single localized defect in the normal-phase region for graphene (cf. Ref.~\cite{bolmatov_graphene_sns}), it is possible to derive analytical expressions for the ABS-energy levels when the radius of the defect is much much smaller than $L$. here, it was clearly seen that the Andreev levels are modified by the impurity. For our three-band semimetal, it may or may not be possible to do so, because it all depends whether we can obtain approximate forms of the resulting wavefunctions and, in the end, whether we can find the solutions from the determinant derived from the matching boundary conditions.

In materials hosting pseudospin-1-type band-crossings, typical Fermi energy $E_F$ can be typically $\lesssim 40$ meV (see, e.g., Ref.~\cite{prb108035428}). Effective barrier strength ($V_0$) of $\sim 500$---$1000$ meV and barrier width measuring $L \sim 20$---$ 50$ nm can now be routinely achieved in realistic experiments \cite{geim-nature}. These junctions, therefore, meet our criteria, viz. $V_0 \gg E_F  $ and $(V_0-E_F) \gg E_F $.
In contemporary experiments, Josephson junctions are routinely fabricates in laboratories, e.g., using graphene \cite{jj-expt-graphene} (for 2d cases) and Weyl semimetals \cite{Ohtomo_2022} (for 3d cases). To carry out such transport measurements on pseudospin-1 semimetals, one has to use an appropriate material hosting triple-point quasiparticles.
The oscillatory behaviour of the Josephson current, arising from a variation in $V_0$ (see, for example, Fig.~\ref{figvl}), shows that we can study Fabry-P\'erot interference using our junction configurations \cite{soori}. One can also create a two-junction quantum interferometer \cite{luca}, and obtain an interference pattern analogous to what is seen in optics for the Young’s double-slit experiment.

In the future, it will be worthwhile to investigate the Josephson effects in twofold and multifold semimetals with nonlinear-dispersion directions \cite{debabrata-krish, debabrata, Deng2020, ips-aritra, ips_jns, Boettcher, ips-qbt-sc}, by considering the propagation of the quasiparticles/quasiholes along such an axis. 
A nontrivial topology can be exhibited when we consider an order parameter comprising multiple components that transform under the same irreducible representation, but not for the case of a single order parameter corresponding to s-wave pairing. This has been discussed in a recent preprint \cite{topo-jj}. The topology of the gap function is manifested through
its Berry-phase structure, which will have its fingerprints in the Josephson current. In the future, we would like to investigate this aspect as well.
Other interesting avenues to explore will be to consider S-N-S or S-B-S junctions comprising higher-angular-momentum pairing channels (e.g., d-wave symmetric pairing channel \cite{igor2}) and/or for FFLO order parameters \cite{emil_jj_WSM, debabrata}.

\section*{Acknowledgments}

This research, leading to the results reported, has received funding from the European Union's Horizon 2020 research and innovation programme under the Marie Skłodowska-Curie grant agreement number 754340.

%
%

\bibliography{biblio1abs}

\end{document}